# An Adaptive Agent Oriented Software Architecture[*]

Babak Hodjat        Christopher J. Savoie        Makoto Amamiya

Department of Intelligent Systems
Graduate School of Information Science and Electrical Engineering
Kyushu University
6-1 Kasugakoen, Kasuga-shi
Fukuoka 816, Japan
http://www_al.is.kyushu-u.ac.jp/~bobby/index.html

**Abstract.** A new approach to software design based on an agent-oriented architecture is presented. Unlike current research, we consider software to be designed and implemented with this methodology in mind. In this approach agents are considered adaptively communicating concurrent modules which are divided into a white box module responsible for the communications and learning, and a black box which is the independent specialized processes of the agent. A distributed Learning policy is also introduced for adaptability.

**Topics:** Agent Architectures, Agents Theories.

**Keywords**: Agent-oriented systems, Multi-Agent Software architectures, Distributed Learning,

---

# 1. Introduction

The classic view taken with respect to *Agent Oriented Systems* is to consider each agent an autonomous individual the internals of which are not known and that conforms to a certain standard of communications and/or social laws with regard to other agents [5]. Architectures viewing agents as such have had to introduce special purpose agents (e.g., broker agents, planner agents, interface agents…) to shape the structure into a unified entity desirable to the user [2]. The intelligent behavior of these key agents, with all their complexities, would be vital to the performance of the whole system.

Another trend in this view is to give the possibility to the agents to query each other's internal knowledge and states through the communications protocols, while at the same time conserving the black-box view of the agents. Inevitably, defining and controlling such issues as conflicts of interest between agents, honesty, helpfulness, and gullibility, have had to be taken into account and dealt with [4]. The most important aspect of this dominant view is that agent architectures are considered to be unifiers of pre-written, separate modules (heterogeneity) [3]. Each of these modules was probably designed without having this higher structure in mind, and is completely different (be it in the code, the machine it is implemented upon, the designer, or the purpose of design). Agent-based Software Engineering was originally invented to facilitate the creation of software able to interpolate in such settings and application programs were written as software agents [6]. On the other hand, methodologies dealing with the internal design of agents tend to view them primarily as intelligent, decision-making beings. In these methodologies, techniques in Artificial Intelligence, Natural Language Processing, machine learning, and adaptive behavior seem to overshadow the agent's architecture, in many cases undermining the main purpose of the agent [7][13].

For instance, one can view agents as reinforcement learning agents with a set of tools to be chosen with respect to environmental senses. Such a view, however well suited for agent learning techniques, may not be readily applied to more algorithmic applications, thus misleading one to assume that such applications should not be or could not be implemented as agents.

In this paper, we wish to present an agent-oriented methodology, which can be universally applied to any software design. The Adaptive Agent Oriented Software Architecture (AAOSA) builds upon and extends the widely accepted object oriented approach to system design. The primary difference sited between Agent-oriented and Object oriented programming has been the language of the interface [6]. In this paper we will suggest an approach in which communication between agents can be done independent of language. This language independent communication will still hold as the main difference with the Object oriented methodology. Another aspect that makes agents more attractive to use in software than objects is their quality of volition. Using AI techniques, adaptive agents are able to judge their results, then modify their behavior (and thus their internal structure) to improve their perceived fitness. First we will clarify our definition of agents, which is somewhat relaxed with respect to the classic definitions. Then the steps by which software should be designed using

AAOSA methodology are described. Some suggestions as to how adaptive learning and communication language independence can be achieved are briefly presented next. To clarify the AAOSA methodology, we present an example application in the form of a simple multimodal map program and show some of the resulting features.

## 2. Our Definition of Agents

Our definition of agents is more in line with the ones given by [12] and [2], and we classify our agents as having the following properties [5]: reactivity, autonomy, temporal continuity, communicative capabilities, team orientation, mobility, learning (adaptive), and flexibility. The resulting multi-agent system we have in mind is a partially connected one [5]. A direct communication specification-sharing approach is taken here to enhance collaboration. Instead of using assisted coordination, in which agents rely on special system programs (facilitators) to achieve coordination [2], in our approach new agents supply other agents with information about their capabilities and needs. To have a working system from the beginning, the designers preprogram this information at startup. This approach is more efficient because it decreases the amount of communication that must take place, and does not rely on the existence, capabilities, or biases of any other program [6].

Adaptive agents are adaptively communicating, concurrent modules. The modules therefore consist of three main parts: A communications unit, a reward unit, and a specialized processing unit. The first two units we will call the white box and the third the black box parts of an agent (Figure 1). The main responsibilities of each unit follow:

*The communications unit:* This unit facilitates the communicative functions of the agent and has the following sub-systems:
- *Input of received communication items:* These items may be in a standard agent communication language such as KQML. Later in this paper we will see that only a small subset is needed here.
- *Interpreting the input:* Decides whether the process unit will need or be able to process certain input, or whether it should be forwarded to another agent (or agents). Note that it is possible to send one request to more than one agent, thus creating competition among agents.
- *Interpretation Policy:* (e.g., a table) Determines what should be done to the input. This policy could be improved with respect to the feedback received for each interpretation from the rewards unit. Some preset policy is always desirable to make the system functional from the beginning. In the case of a system reset, the agent will revert to the basic hard-coded startup information. The interpretation policy is therefore comprised of a preset knowledge base, and a number of learned knowledge bases acquired on a per-user basis. A *learning* module will be responsible for conflict resolutions in knowledge base entries with regard to feedback received on the process of past requests. Past requests and what was done with them are also stored until in anticipation of their feedback.

- *Address-Book:* keeps an address list of other agents known to be useful to this agent, or to agents known to be able to process input that can not be processed by this agent. Requests to other agents may occur when:
  - ➢ The agent has received a request it does not know how to handle,
  - ➢ The agent has processed a request and a number of new requests have been generated as a result.

  This implies that every agent have an address and there be a special name server unit present in every system to provide agents with their unique addresses (so that new agents can be introduced to the system at run time). This address list should be dynamic, and therefore adaptive. It may be limited; it may also contain information on agents that normally send their requests to this agent. In many cases the Address-book could be taken as an extension of the Interpretation Policy and therefore implemented as a single module.
- *Output:* Responsible for sending requests or outputs to appropriate agents, using the Address-book. A confidence factor could be added to the output based on the interpretations made to resolve the input request or to redirect it. We shall see later in the paper that this could be used when choosing from suggestions made by competing agents by output agents.

*The rewards unit:* Two kinds of rewards are processed by this module: outgoing and incoming. An agent is responsible for distributing and propagating rewards being fed back to it[*]. This unit will determine what portion of the incoming reward it deserves and how much should be propagated to requesting agents. The interpreter will update its interpretation policy using this feedback. The rewards will also serve as feedback to the Address-book unit, helping it adapt to the needs and specifications of other agents. The process unit could also make use of this feedback.

The rewards may not be the direct quantification of user states and in most cases will be interpretations of user actions made by an agent responsible for that. We will further clarify this point later in the paper.

*The process unit:* This unit is considered a black box by our methodology. The designer can use whatever method it deems more suitable to implement the processes unique to the requirements of this agent. The only constraints being that the process unit is limited to the facilities provided by the communications unit for its communications with other agents. The process unit also may use the rewards unit to adapt its behavior with regard to the system. Note that each agent may well have interactions outside of the agent community. Agents responsible for user I/O are an example of such interactions. These agents generally generate requests or initiate reward propagation in the community, or simply output results.

The white box module can easily be added to each program module as a *transducer*. According to definition [6] the transducer mediates between the existing program (the process unit) and other agents. The advantage of using a transducer is that it requires no knowledge of the program other than its communication behavior.

---

[*] A special purpose agent is responsible for the interpretation of user input as feedback to individual user requests. This agent will then initiate the reward propagation process.

We mentioned the process unit as being able to conduct non-agent I/O. It is easy to consider I/O recipients (e.g. files or humans) as agents and make the program redirect its non-agent I/O through its transducer. Other approaches to agentification (wrapper and rewriting) are discussed in [6].

## 3. Software design

The software as a whole should be thought of as a society, striving to accomplish a set of requests. The input requests are therefore propagated, processed by agent modules that may in turn create requests to other agents. Again, it is up to the designers to break down the system, as they feel suitable. Hierarchies of agents are possible and agents can be designed to be responsible for the minutest processes in the system. It is advisable that each agent be kept simple in its responsibilities and be limited in the decisions it needs to make to enhance its learning abilities. The overhead of the required units (the white box) should be taken into consideration.

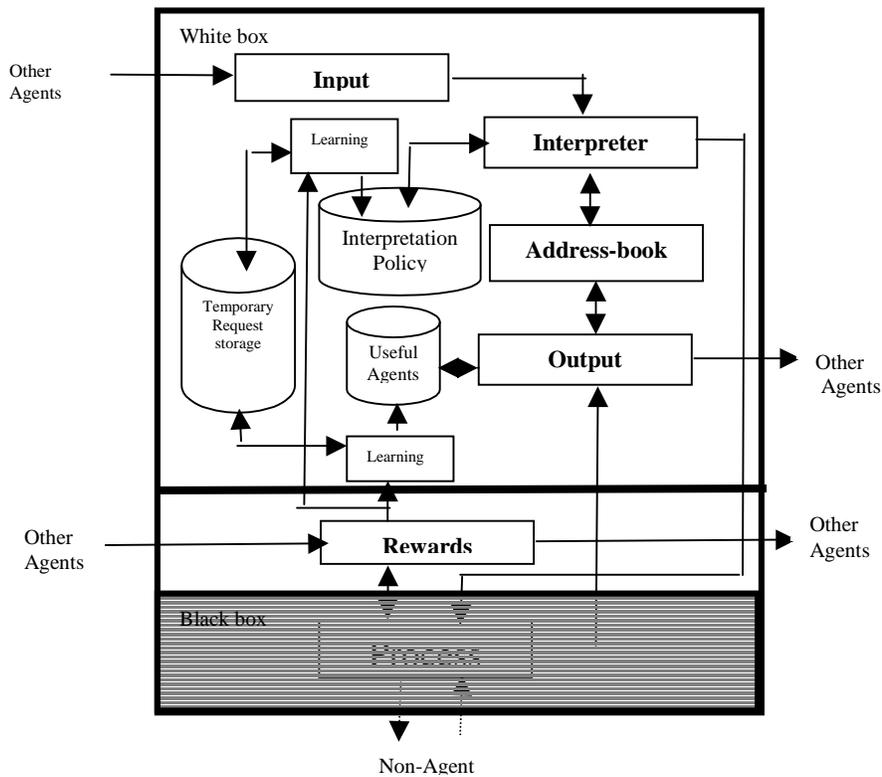

**Fig. 1.** Each agent is comprised of a black box section (specialties) and a white box section (communications).

Agents can be replaced at run-time with other more complete agents. The replacement can even be a hierarchy or network of new agents breaking down the responsibilities of their predecessor. This feature provides for the incremental design and evaluation of software.

We recommend each agent's responsibilities and actions to be clearly defined at design time. As stated in the previous section, many aspects of the white box units should also be preset for each agent according to its definition. To have a working system from the beginning, it is also necessary to define the preliminary communication links between the agents at design time. It should be noted that these communications might change through time, for instance in the case of the introduction of newer agents. Thus, the most important phase in the design of software with this methodology will be to determine the participating agents and their capabilities, although the precise details of how they will eventually communicate with each other may not be known at design time.

There are a number of ways by which the designer can limit the changes that his design may undergo in the future or at run time to guarantee a certain degree of functionality for the design. Introduction of new agents could be constrained in the Address Book of critical agents stopping them from passing requests to alien agents. Certain special purpose agents such as the input or output agents could also serve to limit unwanted future changes to the system.

**3.1. Special purpose agents**
Some special purpose agents may be used depending on the application, for example, agents directly involved with input and output, or an agents which interprets the actions of the user as different levels of reward to different system output[*] (Figure 2)

*Input Agents*
Inputs to the system may be from more than one source. In such systems, one, or a network of special purpose input agents should be considered, which:
- Unify inputs from different sources into one request, and/or
- Determine commands that all fall into a single request set.

For example if the user's Natural Language (NL) input is: "Information on this" and the mouse is then pointed at an item on the display, these two inputs should be unified as a single request. Interactive input would also require the input agents to determine the end of an input stream. For instance in NL input a phrase (e.g., Do! or Please!) or a relatively long pause, could determine the end of the stream. A different policy here would be to interpret the input in real-time and redirect it to the appropriate agent. As seen in figure 2, agents can redirect the input back to the input agents once this data is no longer relevant to the responding agent.

*Output Agents*
This special purpose agent decides which response suggested by various agents should be actuated, thus ensuring competition between agents. The decision may be made based on a combination of different criteria and may depend on a specific request. The criteria may be speed, confidence, or other checks that could in turn be made by quality assurance agents. After the final choice has been made, the output agent will ask the specific suggesting agent to actualize its suggestion, or the decision may be redirected to actuator agents. The output agents may be more than one, thus breaking the decision making process into a hierarchy. For instance, competing agents could have an output agent decide between them and the output agents in turn could have a higher-level output agent. Safety mechanisms to ensure the requests will not be stuck in cycles of infinite deadlock loops between the agents could be another responsibility of the output agents. Special purpose safeguard agents could also be used.

---
[*] One of these agents may be the reward agent itself, thus tuning itself with the user.

*Feedback Agents*
Any adaptive system needs a reward feedback that tells it how far from the optimum its responses have been. This reward could be explicitly input to the system, or implicitly judged from input responses by the system itself. In the case of implicit rewarding, an agent could be responsible for the interpretation of the input behavior and translating it into rewards. The criteria that could be used depend on the system. For instance in an interactive software application a repeat of a command, remarks indicating satisfaction or dissatisfaction, user pause between requests or other such input could be translated into rewards to different output. The feedback agents could also be more than one depending on the different judgement criteria and a hyper-structure [8] or hierarchy might be designed to create the rewards. One way of propagating the feedback reward through the system would be to be aware of the different output and to pass the interpreted reward to the final (output) layer, which will, in turn, pass it back to the previous agents involved in that particular output.

A name server unit is also required to make possible the dynamic introduction of new agents to the system. Each new agent will have to obtain its name (or address) from this name server so that conflicts do not occur and so agents can be referred to throughout the system. Input requests (commands) to the system should be tagged with the user-id that has issued them because interpretation is done on a per-user basis. Reward fed back to the system should also incorporate the request-id to which the reward belongs.

## 4. Communication Language

In Agent-Based Software Engineering [6], Agents receive and reply to requests for services and information using a declarative knowledge representation language KIF (Knowledge Interchange Format), a communication language KQML (Knowledge Query and Manipulation Language) and a library of formal ontologies defining the vocabulary of various domains. KQML [4] is a superset of what is needed as a communication language between the AAOSA agents and may well be used effectively. The stress on adaptability eliminates the need for elaborate languages and even a simple message passing protocol between the agents should be sufficient.
The main request string may be made of different types of information (e.g., Character strings, voice patterns, images, etc…). Various standard information may be passed along with the main request string including: The originator agent, sender agent, the user initiating this request, request id, and/or a time stamp. The request string itself could be comprised of any item of information ranging from natural language requests from the user, to specialized inter-agent messages. Introductory messages sent by new agents (introduced to the system at run time), could also be incorporated in the request string, or sent under preset standards such as those provided by KQML. Other specialized (and possibly standardized) information that should be passed includes the reward propagation data.

## 5. An Example: A multimodal map [1]

Multiple input modalities may be combined to produce more natural user interfaces. To illustrate this technique [1] presents a prototype map-based application for a travel-planning domain. The application is distinguished by a synergistic combination of handwriting, gesture and speech modalities; access to existing data sources including the World Wide Web; and a mobile handheld interface. To implement the described application, a distributed network of heterogeneous software agents was augmented by appropriate functionality for developing synergistic multimodal applications. We will consider a simplified subset of this example to show the differences of the two approaches. A map of an area is presented to the user and she is expected to give view port requests (e.g., shifting the map or magnification), or request information on different locations on the map. For example, a user drawing an arrow on the map may want the map to shift to one side. On the other hand the same arrow followed by a natural language request such as: "Tell me about this hotel." May have to be interpreted differently.

[Cheyer et al 96] use the Open Agent Architecture (OAA) [2] as a basis for their design. In this approach, based on a "federation architecture" [9], the software is comprised of a hierarchy of facilitators and agents. The facilitators are responsible for the coordination of the agents under them so that any agent wanting to communicate with any other agent in the system must go through a hierarchy of facilitators (starting from the one directly responsible for it). Each agent, upon introduction to the system, provides the facilitator above it with information on its capabilities (Figure 3). No explicit provision is given for learning.

An example design based on AAOSA is shown in figure 4. It must be noted here that the design shown in figure 4 is not rigid and communication paths may change through time with the agents adapting to different input requests. The NLP and pointer input agents determine the end of their respective inputs and pass them on to the input regulator. This agent in turn determines whether these requests are related or not. It then passes it down to the agent it considers more relevant to the request. The output agents hierarchically sift the outputs suggested by the shifting, magnification, hotels, restaurants and general information agents. Note that combinations of these output suggestions could also be chosen for actuation. The feedback agent provides the system with rewards interpreted from user input. Some of the differences in the two designs are given below:

- The AAOSA design is much more distributed and modular by nature and many of the processes concentrated in the facilitator agents in figure 3 are partitioned and simplified in figure 4.
- AAOSA is more of a network or hyper-structure [8] of process modules as opposed to the hierarchical tree like architecture in the OAA design.
- AI behavior such as natural language processing and machine learning are incorporated on the architecture rather than introduced as new agents (as is the case with the natural language macro agent in figure 3).

It must be stressed that AAOSA like architectures could be achieved with an OAA if we take each OAA facilitator and its macro agents as one agent and add learning capabilities to each facilitator. Another point worth mentioning is that agents in OAA are usually pre-programmed applications linked together through facilitators. The designers have a lesser say over the software architecture as a whole because they are forced to use what has already been designed, possibly without the new higher-level framework in mind.

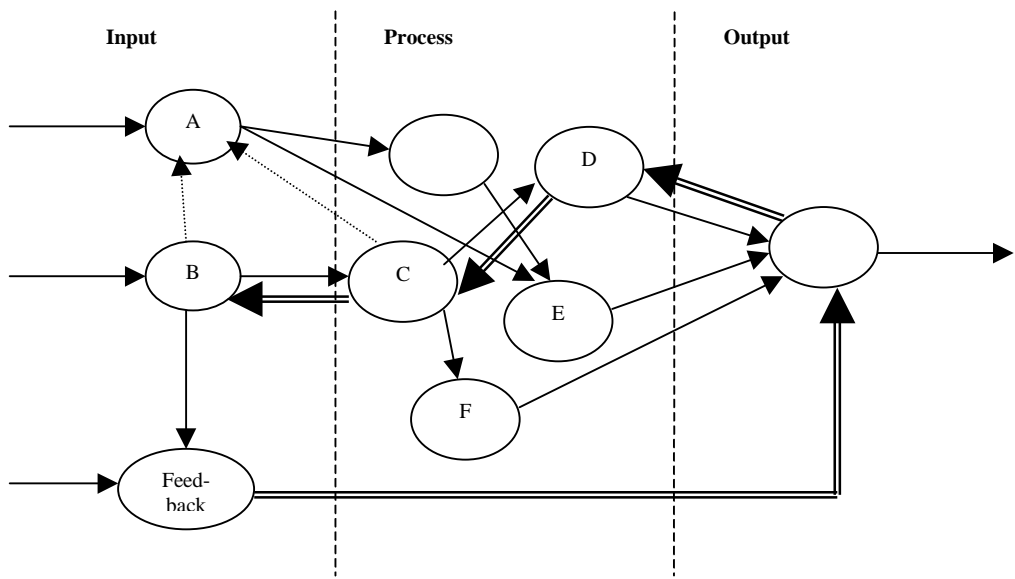

**Fig. 2.** Example of agent-oriented software architecture. Input agent B may redirect input that does not belong to it to agent A (dotted arrow). This redirection may even happen in later stages (e.g., from C to A). Output suggestions from agents D, E, and F are considered in the output agent and one (in this case D's) is chosen for actuation. The feedback agent uses input directives or indirect behavioral assumptions to calculate a numerical representation of the reward for each output. This reward is then fed back to the agents (e.g., double line arrows).

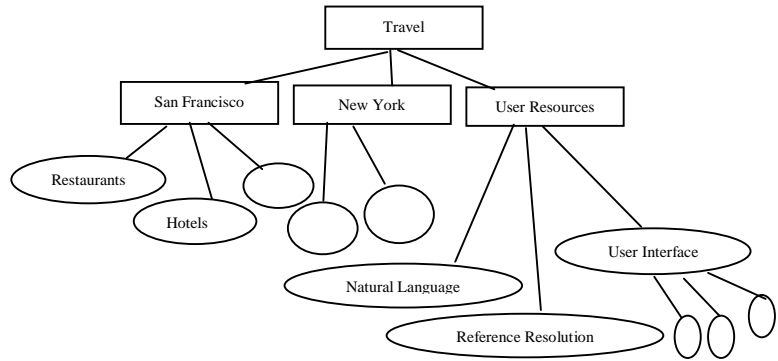

**Fig. 3.** A structural view of the multimodal map example as designed using OAA in [1]. Boxes represent Facilitators, ellipses represent macro agents and circles stand for modality agents.

## 6. Interpretation policy

Each agent matches the input pattern with its stored key patterns and finds the closest match and a degree of confidence for it. If this degree of confidence is lower than a threshold, the agent will forward the request to other agents for whom it is confident of their ability to process the request. If no other agent is known with these specifications, either a random action (weighted by the confidence on that action) is chosen, or the request is forwarded to another agent.

In the simplest case the pattern matching process is comprised of checking the presence or absence of certain segments (e.g., words) to determine the course of action that needs to be taken. In more complicated forms patterns should be discovered in the context of the input (e.g., grammar or semantics). This is one reason why if each agent is kept simple in the range of the decisions it needs to make based on its input, the matching and learning process is simplified. A simple pattern matching may be enough to determine the course of action needed. For instance the occurrence of a pointer drag or word patterns including such phrases as "go", or "shift" could cause the map agent in our example (Figure 4) to redirect the request to the shifting agent. Whereas words such as "bigger", "smaller", "magnify", "I can't see" may cause it to send the request to the magnification agent (Figure 5).

Techniques such as those given in [10] could help sort the patterns according to their information value. The nature of the information depends on the application and the agent specializing in it. For example such information as the time between inputs and the loudness or general pattern of the input speech wave could be useful for the feedback agent. The information value of patterns varies depending on the agent (Figure 5).
We will not offer any solutions as to how the interpretation policy of each agent should be stored and updated. Some points that should be taken into consideration while pondering a solution follow:

- It is very important for the interpreter agent to be able to load pre-defined policies at start-up. These are not learned, but hard wired by the engineers. The engineers also determine how much of this initial policy could be undermined through learning.

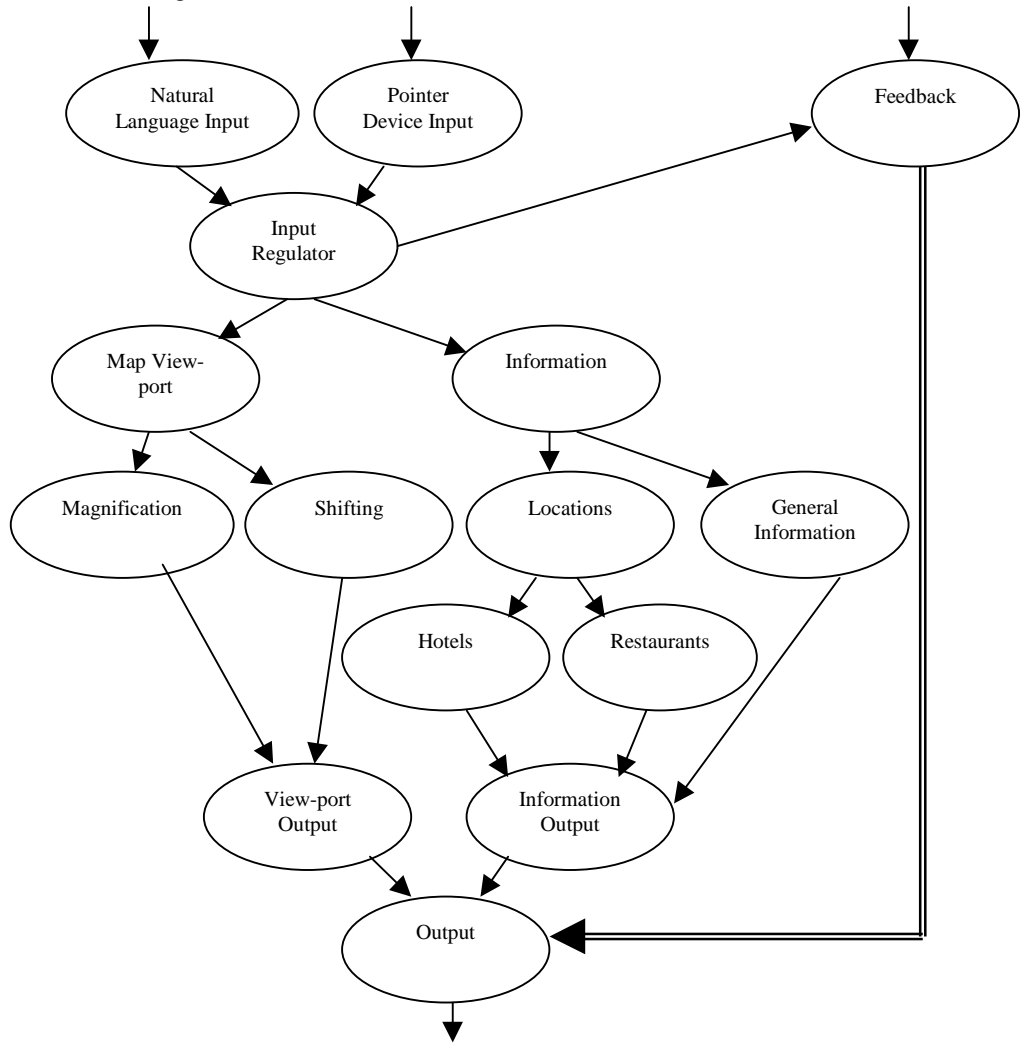

**Fig. 4.** The multimodal map example as designed based on AAOSA.

- It is of equal importance for other agents to be able to contribute to this policy, for instance introducing themselves to the Address-book as possible references.
- The interpretation policy should be dynamic to allow learning of new interpretation rules.
- In many cases the learned policies should be stored on a per-user basis.

## 7. Learning

Adaptability in AAOSA materializes in three forms:
- The ability of the system to accept new agents at run time,
- The ability of each agent to accept unexpected input or requests,
- The ability of each agent to adapt its behavior according to the feedback it receives (i.e., learning).

Some features of a pattern learning algorithm that may be suitable for AAOSA are briefly mentioned in this section.

Each agent upon the input of a new request goes through the following steps. It must be noted again that the choices mentioned here might be either internal processes or other agents to direct the request to.
- Scan input request for stored patterns estimating a confidence value for each match.
- Choose nearest pattern's choice.
- Keep track of patterns being thrown away in the process of matching as low information patterns [10].
- In case of close ties, choose at random between higher confidence options.
- In case of no reliable match, choose at random between all options[*].
- Store request-choice decided upon for adjusting weights and learning until the feedback arrives (delayed reward).

In case of negative reward, patterns in request with highest conflict resolution value should be stored as new decision criteria. These new patterns will be stored according to the user so different users will receive different responses based on their profile in each agent. For example in figure 5 if the input request is slightly changed to: "Shift the *view* to the right", a contradiction will occur. This contradiction could be resolved if the agent identifies the pattern "view" as a higher information value pattern and a new interpretation policy based on the absence or presence of this pattern is conceived.

## 8. Conclusion

Viewing software as a hyper-structure of Agents (i.e., intelligent beings) will result in designs that are much different in structure and modularization. Some of the benefits of this approach are noted here, some of which are also achievable in object oriented design.
- Flexibility: There is no rigid predetermination of valid input requests.
- Parallelism: The independent nature of the agents creates a potentially parallel design approach.
- Multi platform execution: Agents can run and communicate over networks of computers (on the Internet for instance).
- Runtime addition of new agents and thus incremental development of software.

---

[*] Random functions may be weighted according to confidence.

- Software additions by different designers: Different designers can introduce different agents to compete in the software, making this design methodology attractive for commercial applications.

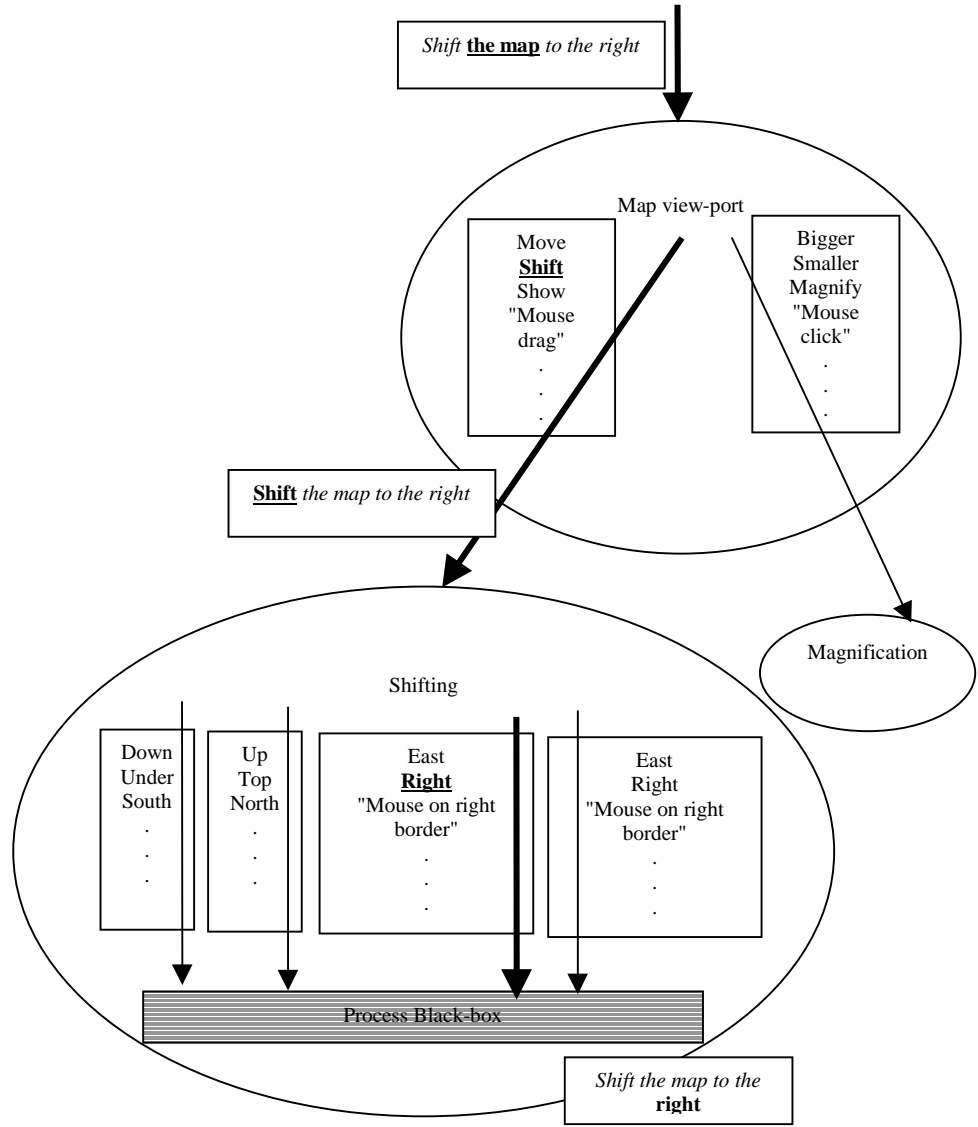

**Fig. 5.** Each agent needs to identify a small subset of the information in the request and act upon that. This is also an example of distributed Natural Language Processing. Low information (throwaway) patterns (shown in *italic*) vary depending on the agent.

- Reusability of agents.
- Incremental design and evaluation.
- Learning and Intelligence: The distributed nature of learning introduced in this paper suggests a powerful adaptive software design that potentially breaks down an application to a hyper-structure of simple learning modules [8]. Another AI technique that could readily be incorporated into the agents is *artificial evolution* [11]. The mere presence of a reward for each agent makes the introduction of death (removal of an agent from the software) possible. This will make way for other agents, perhaps with better learning techniques, to take over. There will also inevitably be numerous variables to be fine-tuned for each agent. These variables may be thought of as the agent's genes and optimized through this evolutionary process.